\documentclass[letter,scriptaddress,twocolumn, prl,showkeys]{revtex4}

	\usepackage{amsmath}
	\usepackage{makeidx}
	\usepackage{amsfonts}
	\usepackage[ansinew]{inputenc}
	\usepackage[usenames,dvipsnames]{pstricks}
	\usepackage{subfigure}
	\usepackage{epsfig}
	\usepackage{pst-grad} 
	\usepackage{pst-plot} 
	\usepackage[colorlinks,hyperindex]{hyperref}
    \usepackage{graphicx}
    \usepackage{epstopdf}

    \DeclareGraphicsExtensions{.pdf,.png,.jpg,.svg,.eps}

	\hypersetup
	{
		colorlinks,%
		citecolor=black,%
		linkcolor=black,%
		urlcolor=black,%
	}

\makeindex

\begin{document}

\title{Link Prediction via Matrix Completion}

\author{Ratha Pech$^1$, Hao Dong$^{1,2,*}$, Liming Pan$^1$, Hong Cheng$^3$, Zhou Tao$^{1,2,}$}

\email{haodong@uestc.edu.cn \\
       zhutou@ustc.edu}
\affiliation{$^1$ CompleX Lab, University of Electronic Science and Technology of China, Chengdu 611731, People's Republic of China}
\affiliation{$^2$ Big Data Research Center, University of Electronic Science and Technology of China, Chengdu 611731, People's Republic of China}
\affiliation{$^3$ Center for Robotics, University of Electronic Science and Technology of China, Chengdu 611731, People's Republic of China}


\begin{abstract}
  Inspired by practical importance of social networks, economic networks, biological networks and so on, studies on large and complex networks have attracted a surge of attentions in the recent years. Link prediction is a fundamental issue to understand the mechanisms by which new links are added to the networks. We introduce the method of robust principal component analysis (robust PCA) into link prediction, and estimate the missing entries of the adjacency matrix. On one hand, our algorithm is based on the sparsity and low rank property of the matrix, on the other hand, it also performs very well when the network is dense. This is because a relatively dense real network is also sparse in comparison to the complete graph. According to extensive experiments on real networks from disparate fields, when the target network is connected and sufficiently dense, whatever it is weighted or unweighted, our method is demonstrated to be very effective and with prediction accuracy being considerably improved comparing with many state-of-the-art algorithms.
\end{abstract}


\maketitle

\section{Introduction}

In the past decade, the rapidly expanding of studies on complex networks has brought together different disciplines including physics, mathematics, computer science, sociology, economics, biology and so on \cite{newman2010networks,albert2002statistical}. The theory of complex networks provides us novel insights for understanding the real-world linking patterns. The real-world linked datasets are usually dynamically changing and subjected to unobservability. On one hand, the datasets are growing and changing over time through the increment of new links \cite{holme2012temporal}. On the other hand, the missing or unobservable entries extensively exist in the datasets \cite{guimera2009missing}. Therefore, predicting missing links is of great importance to signifying the newly appeared and unobserved relations between data entries.

Link prediction problem is essentially involved with the knowledge discovery and topology remodeling for large volumes of dynamic and noisy datasets \cite{getoor2005link}, which also aims at uncovering to what extent the evolution of networks can be modeled and analyzed according to the intrinsic features and structures of the network itself \cite{wang2012evaluating}. So far it has been generally accepted as a fundamental paradigm not only in physics but also in bioinformatics, sociology, statistics and computer science. For instance, in biological science, predicting the interactions between proteins in protein interaction networks, creatures in food-web networks and biochemical outcomes in  betabolic networks can help one in checking firstly the most likely existing links rather than checking all the possible interactions \cite{clauset2008hierarchical}. In recommender systems, target networks are user-item bipartite networks, and link prediction is to recommend items to users, so as to help the users effectively and efficiently surf the products and consequently improve the sales \cite{lu2012recommender}. For social media, such as Facebook, Twitter, Weibo, Tecent WeChat and so on, link prediction aims at recommending friends to users thus enhance their loyalties to the sites \cite{liben2007link}.

A lot of effort has been made to solve the link prediction problem \cite{lu2011link,lu2015toward,liu2015local,scholz2014link,berlusconi2016link,zhu2016link,zhang2016measuring,barzel2013network} and most of them are based on similarity between vertex pairs since these algorithms are designed according to the fact that similar vertices are more likely to connect to each other. These algorithms are called similarity-based algorithms. Roughly, the similarity indices can be classified into three categories \cite{lu2011link}, e.g., local \cite{newman2001clustering,zhou2009predicting}, global \cite{katz1953new,brin2012reprint} and quasi-local \cite{lu2009similarity,liu2010link} indices. The most popular methods are the local ones because they are simple and applicable for very large-scale networks. Although they are computational efficient, the local similarity-based link prediction algorithms are sometimes less accurate.

The global topological information can be exploited through the adjacency matrix, where the nonzero entries denote the connections between vertices, while missing links and nonexisting links are both denoted by zero entries. In most cases, a very small fraction of zero entries (called hidden nonzero entries or hidden entries) represent the missing links and the rest (called null entries) represent the nonexisting links. Essentially speaking, a link prediction algorithm aims at recovering the hidden non-zero entries from the real null entries according to all the non-zero entries in the adjacency matrix. However, for a real-world network, the adjacency matrix is usually very sparse (i.e., most of its entries are zeros), providing highly limited information. How to precisely predict the missing links based on the sparse information is a challenging issue. Recently, the principal component analysis (PCA) shows that a matrix in which certain entries are missing or corrupted can be successfully reconstructed as long as the original matrix has the low-rank property, where the rank of a matrix is defined as the total number of the linearly independent columns or rows.

In this work, we introduce the robust principal component analysis (robust PCA) method into link prediction and design a novel global information based prediction algorithm based upon low rank and sparsity property of the adjacency matrix. We then reconstruct a network that is close to the original network and accordingly identify missing links by discovering the matrix with minimum nuclear norm which fits the training data. It is shown that when the target network is connected and sufficiently dense, we can find out the missing links with much higher accuracy comparing to the state-of-the-art algorithms.

\section{Method}
An undirected network consists of a set of vertices $V$ and a set of links $E$. We do not consider multiple links and self-connections. Suppose we have an observed network represented by adjacency matrix $\mathbf{A}\in \mathbb{R}^{n\times n}$, which is a snapshot or a subset of an original network $\mathbf{G^*}$. The set of links in $\mathbf{A}$ and $\mathbf{G^*}$ are denoted by $E^T$ and $E$, respectively. Denote the rest of links in $E$ as $E^P$, namely $E=E^T\cup E^P$ and $E^T\cap E^P=\emptyset $. Then $E^T$ is the training set for learning and prediction and $E^P$ is the probe set for verifying the prediction accuracy. Without loss of generality, in the experiment we dynamically take $80\%, 85\%, 90\%$, and $95\%$ of all links in $\mathbf{G}^*$ as the training set and the rest as the probe sets, respectively.
{\small{
\begin{table*}
\caption{The topology of the twelve real networks. $|V|$ and $|E|$ are the number of vertex and link, respectively. $C$, $r$ and $\langle k \rangle$ are cluster coefficient, assortative coefficient and average degree, respectively. $H = \frac{\langle k^2 \rangle}{\langle k \rangle^2}$ is the degree heterogeneity of network computed. R, $\tau$ and D are the rank of adjacency matrix, ratio between rank and dimension and network density, respectively.}
\centering
\begin{tabular}{ l  l  l l l l l l l l}
\hline
\hline   \\[-2.5ex]                     
Networks & $|V|$ \ & $|E|$ \ & $C$ \ & $r$ \ &$\langle k \rangle$ \ &$H$ \ & R \ & $\tau$ \ & D \\ 
\hline
Jazz            & 198 & 2742 & 0.618 & 0.02 & 27.697 & 1.395 & 198 & 1.000 & 0.1406\\
Yeast           & 2375 & 11693 & 0.306 & 0.45 & 9.850  &  3.474 & 1816 & 0.765 & 0.0042 \\
Political blogs & 1222 & 19021 & 0.320 & -0.22 & 27.355 &  2.970  & 1093 & 0.894 & 0.0224 \\
Hamster         & 1858 & 12534 & 0.141 & -0.09 & 13.492 &  3.361  & 1221 & 0.657 & 0.0073 \\
Router          & 5022 & 6258 & 0.012 & -0.14 & 2.493 &  5.502  & 3054 & 0.608 & 0.0005 \\
Food web 1      & 128 &	2106 & 0.335 & -0.10 & 32.422 & 1.237   & 124 & 0.969 & 0.2553\\
World trade     & 80 & 875 & 0.753 & -0.39 & 21.875 & 1.558   & 79 & 0.988 & 0.2769 \\
Contact         & 264 & 2108 & 0.658 & -0.48 & 15.970 & 3.546   & 82 & 0.311 & 0.0607 \\ [1.2ex]

USAir           & 332 & 2126 & 0.013 & -0.21 & 12.807 & 4.915  & 274 & 0.825 & 0.0387 \\
C.elegans       & 306 & 2148 & 0.647 &  -0.16 & 14.039 & 4.642  & 282 & 0.922 & 0.0460 \\
Food web 2             & 69 & 880 & 0.067 & -0.30 & 25.507 & 7.972  & 66 & 0.957 & 0.3751 \\
Football        & 35   & 118   & 0.353 & -0.18 & 6.743 &  1.608  & 35 & 1.000 & 0.1983 \\ [0.5ex]       

\hline 
\end{tabular}
\label{table:topology}
\end{table*}
}}

\begin{figure}[ht]
\centering
\includegraphics[width=0.45\textwidth]{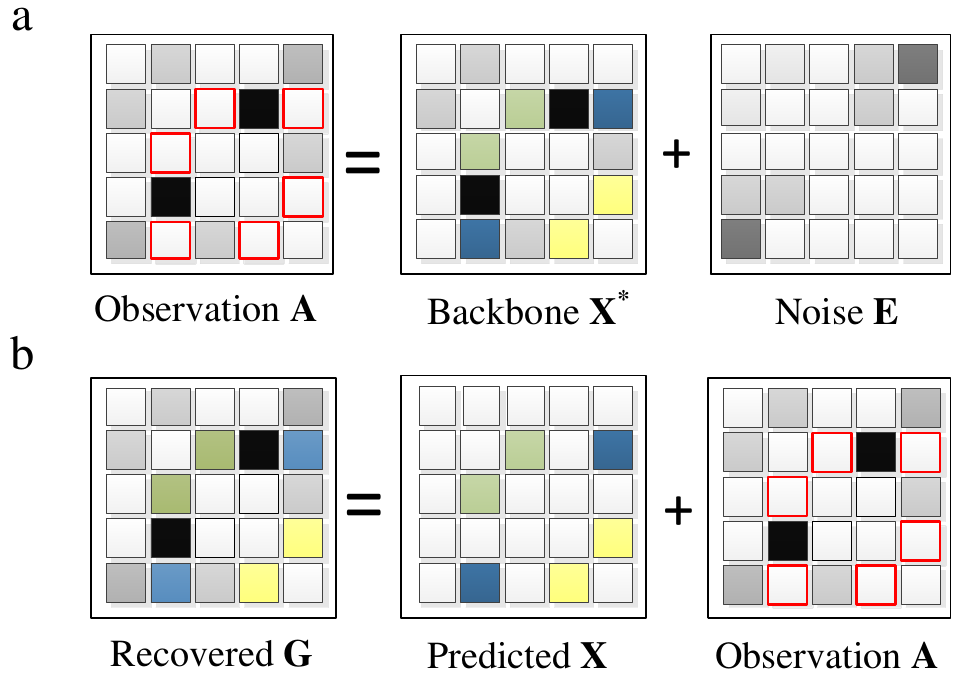}
\caption{(a) The relationship between the observed network $\mathbf{A}$, the corresponding backbone network $\mathbf{X^*}$ containing newly appearing links and some existing links and the noise $\mathbf{E}$ containing spurious links in observed network $\mathbf{A}$. (b) The relationship between the recovered network $\mathbf G$, the predicted network $\mathbf{X}$ containing only newly appearing links and the observed network $\mathbf{A}$. The white colors represent real null entries with value zero, the white colors with red frames represent the missing or likely existing links (values are also zeros), while the others entries with colors are of values greater than zero.}
\label{fig_lr_matrix}
\end{figure}

The objective of link prediction is to find out the missing links of the original network $\mathbf{G}^*$, that is to recover a network $\mathbf{G}$ (it is worth noting that it is generally intractable to recover exactly $\mathbf{G}^*$), which is sufficiently close to $\mathbf{G^*}$, based on the observed entries of $\mathbf{A}$. Assume that (i) $\mathbf{X^*}\in \mathbb{R}^{n\times n}$ conveys the pattern how the network evolves (how new links are added and some old links are eliminated) and we call $\mathbf{X^*}$ the backbone network; (ii) $\mathbf{X}$ is the subset of $\mathbf{X}^*$ containing only the new links, which can be obtained by resorting the values of elements corresponding to non-zero entries in $\mathbf{A}$ to be zero. $\mathbf{X^*}$ and $\mathbf{X}$ have real number values. Identifying network $\mathbf{X^*}$ is the crucial intermediate step for recovering the original network and predicting the missing links accordingly. The observed network $\mathbf{A}$ is the only information we can utilize. $\mathbf{X^*}$ can be represented by subtracting an error/noise matrix $\mathbf{E}\in \mathbb{R}^{n\times n}$ from $\mathbf{A}$ and this noise matrix should be much more sparser than either $\mathbf A$ or $\mathbf X^*$.
Therefore, $\mathbf{X^*}$ can be written as

\begin{equation}
    \mathbf{X^*}=\mathbf{A}-\mathbf{E},
\label{eq_A}
\end{equation}
where $\mathbf{E}$ is the noise matrix in which positive entries are the spurious links and negative entries represent the missing links which appear in $\mathbf{X^*}$. The relationship between $\mathbf{G}$, $\mathbf A$, $\mathbf{X}$, $\mathbf{X^*}$ and $\mathbf{E}$ is illustrated in Fig. 1. The recovered network $\mathbf{G}$ is obtained as 
\begin{equation}
{{\mathbf G}\mathrm{{=}}{\mathbf X}+{\mathbf A}}.\\
\label{eq_GXA}
\end{equation}
$\mathbf{X}$ contains only newly appeared links and it is defined as
\begin{equation}
    x_{ij}=\left\{{\begin{array}{l}
    {x^*_{ij}, \ \ \ a_{ij}=0.}\\
    {0, \ \ \ \ \ a_{ij}=1.}
    \end{array}}\right.
\label{eq_NewLink}
\end{equation}

The Principal Component Analysis (PCA) is a prevailing tool for identifying the hidden patterns and relevant information in datasets based on observed information. It can be utilized to obtain $\mathbf{X^*}$ and $\mathbf{E}$ simultaneously by converting the observed network $\mathbf{A}$ into a set of linearly uncorrelated variables called principal components, which captures the backbone network $\mathbf{X^*}$. PCA requires that both $\mathbf{A}$ and $\mathbf{E}$ have the low-rank property, however, in most real networks, $\mathbf{A}$ is usually very sparse resulting in the dissatisfaction of the low-rank property. Therefore, a more robust matrix completion approach against high-dimensional noise $\mathbf{E}$ is required for link prediction in real complex networks. Hence, we apply the robust principal component analysis (robust PCA) in the matrix completion for link prediction.

Mathematically, according to the theory of robust PCA, recovering matrix $\mathbf{X^*}$ can be transformed into the following optimization problem:
\begin{equation}
    \min_{\mathbf{X^*,E}} \mathrm{rank(\mathbf{X^*})}+\gamma||\mathbf{E}||_0 \ \ \ \mathrm{s.t.} \ \ \ \mathbf{X^*=A-E},
\label{eq_rpca}
\end{equation}
where $\mathrm{rank(\mathbf{X^*})}$ denotes the rank of matrix $\mathbf{X^*}$, the operator $||.||_0$ is the $l_0$-norm (i.e., the number of nonzero entries of a matrix), and $\gamma$ is the parameter balancing these two terms. Normally, a precise solution of $\mathbf{X^*}$ guarantees that $\mathbf G=\mathbf G^*$, which means the precise solution of $\mathbf{X^*}$ can be used to perfectly recover the original network. Finding the precise solution of $\mathbf{X^*}$ in Eq. (\ref{eq_rpca}) is a highly nonconvex optimization problem and its complexity is nondeterministic polynomial. However, the approximate solutions can be obtained based on robust PCA \cite{wright2009robustPCA}. Firstly, since a matrix with rank $r$ has exactly $r$ nonzero singular values,  $\mathrm{rank(\mathbf{X^*})}$ is just the number of nonzero singular values of the matrix $\mathbf{X^*}$. Secondly, according to the pioneer works \cite{candes2005decoding,donoho2006most}, the solution of $l_1$-norm is also a sparse solution of $l_0$-norm. Hence, the tightest relaxation of $\mathrm{rank(\mathbf{X^*})}$ and  $l_0$-norm are the nuclear norm and $l_1$-norm, respectively \cite{candes2009exact,wright2009robust,lin2010augmented}. In a word, the relaxed approximate solution of Eq. (\ref{eq_rpca}) can be written as
\begin{equation}
    \min_{\mathbf{X^*,E}} ||\mathbf{X^*}||_*+\lambda||\mathbf{E}||_1 \ \ \ \mathrm{s.t.} \ \ \ \mathbf{X^* = A - E},
\label{eq_lr}
\end{equation}
where $||.||_*$ denotes the nuclear norm (i.e., the sum of singular values) of matrix, $||.||_1$ is $l_1$-norm (i.e., the sum of the absolute values of matrix entries), $\mathbf{E}$ is a sparse matrix (i.e., most of its entries are zeros) and $\lambda$ is the positive weighting parameter balancing the low-rank property and sparsity.

{\small{
\begin{table*}[ht]
\caption{The prediction precision on the eight real unweighted networks in which the probe set contains $10\%$ of total connections.}
\centering
\begin{tabular}{*{8}{l}}

\hline
\hline                        
Networks \ & CN \ & AA \  &RA \  & CAR \ & CAA \ & CRA \ & LR \ \\ [0.5ex]
\hline
Jazz          & 0.502 & 0.521 & 0.533 & 0.514 & 0.525 & 0.552 & \textbf{0.606} \\
Yeast         & 0.139 & 0.159 & 0.256 & 0.138 & 0.143 & 0.158 & \textbf{0.586} \\
Political blogs  & 0.178 & 0.175 & 0.155 & 0.176 & 0.177 & 0.178 & \textbf{0.212} \\
Hamster       & 0.037 & 0.038 & 0.033 & 0.045 & 0.044 & 0.044 & \textbf{0.462} \\
Router        & 0.018 & 0.016 & 0.008 & 0.020 & 0.020 & 0.022 & \textbf{0.113} \\
Food web 1    & 0.070 & 0.072 & 0.068 & 0.069 & 0.069 & 0.072 & \textbf{0.577} \\
World trade   & 0.402 & 0.420 & 0.430 & 0.395 & 0.416 & 0.423 & \textbf{0.457} \\
Contact       & 0.556 & 0.559 & 0.558 & 0.552 & 0.554 & 0.558 & \textbf{0.600} \\[1ex]       
\hline 
\end{tabular}
\label{table:pre_result_unweighted}
\end{table*}
}
}
\begin{figure*}[ht]
    \centering
	\includegraphics[width=0.93\textwidth]{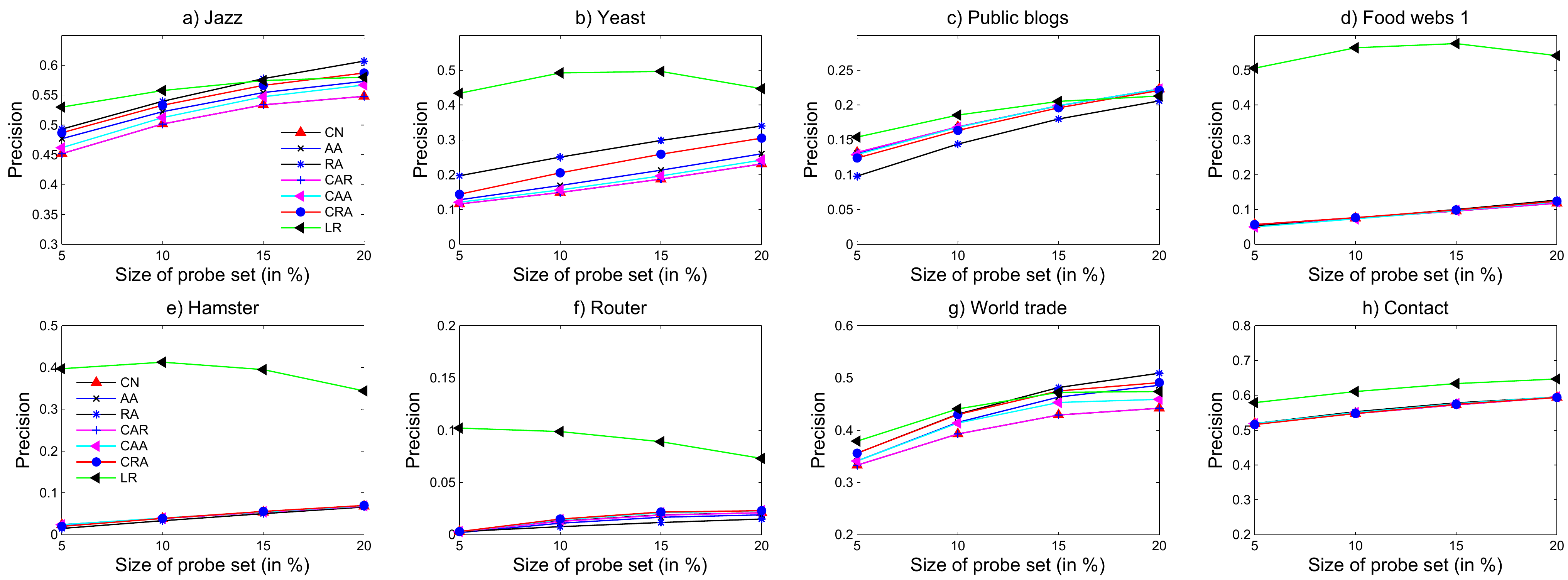}
    \caption{The precision values on the eight real unweighted networks for different sizes of the probe sets.}
\label{fig_probe_uw}
\end{figure*}

The approximate solution of $\mathbf{E}$ and $\mathbf{X^*}$ are, later on, symmetrized as $\mathbf{E}=\mathbf{E}+\mathbf{E}^T$ and $\mathbf{X^*}=\mathbf{X^*}+\mathbf{X^*}^T$, where $\mathbf{M}^{T}$ denotes the transpose of matrix $\mathbf{M}$, (for undirected network and treated as the backbone network generated from the training network $\mathbf{A}$). On the one hand, $\mathbf{X^*}$ contains new links not in $\mathbf{A}$; on the other hand, it also eliminates some possible links in $\mathbf{A}$. After obtaining $\mathbf{X}^*$ we check only the newly appearing links and ignore the observed links in $\mathbf{A}$, as shown in Eq. (\ref{eq_NewLink}), then we merge $\mathbf{X}$ with $\mathbf{A}$ to recover a matrix ${\mathbf G}$ as illustrated in Eq. (\ref{eq_GXA}). This matrix is recovered from the observed data $\mathbf{A}$ through the above procedure, and it is supposed to be close to the original network ${\mathbf G^*}$.

Each pair of vertices (e.g., $x$ and $y$) in ${\mathbf G}$ is bundled with a score $S_{xy}$. If a pair of vertices are linked in both $\mathbf{X}^*$ and $\mathbf{A}$, only the value in $\mathbf{A}$ is used; otherwise, if this pair of vertices are linked only in $\mathbf{X}^*$, it means that this link is a predicted link, thus the value in $\mathbf{X}^*$ is used. Each entry in this score matrix ${\mathbf G}$ denotes the likelihood such that this pair of vertices are connected and it is likely to assign higher values to the missing or the likely existing links than nonexisting links, because the former should have values greater than zeros, while the later should have zero values. It is worth noting that the above approach can also be applied in solving link prediction problem in directed network \cite{zhang2013potential}. Finally, we sort the score of unobserved links in a descending order and select the top $L$ links. In this work,  $L$ is the cardinality of the probe set. We check whether each of these $L$ links really appears in the probe set and record the number of appearing links as  $L_r$. As we set the $L$ as the cardinality of the probe set,  the precision value is also equal to recall value at this point \cite{lu2011link}, as
\begin{equation}
    Pr=L_r/L.
\label{Precision}
\end{equation}

\section{Analysis}
One crucial question is: to what extend we can predict the missing links by utilizing the above matrix completion method? In \cite{candes2009exact}, the authors proved that when the  $m$ observed entries of an $n\times n$ matrix with rank $r$ satisfy the following inequality,
\begin{equation}
    m \geq Cn^{1.2}r\log(n),
\label{eq_lr_inequality}
\end{equation}
where $C$ is a positive constant, one can perfectly recover all entries of the matrix with a very high probability through solving a simple convex optimization problem. However, for the real-world data, the adjacency matrix is very sparse where the order of the number of non-zero entries is normally much less than $n^{1.2}r\log(n)$. Fortunately, for the link prediction problem, it is not required to recover all the non-zero entries of the adjacency matrix, since only a small portion of these zero entries are the missing links and the rest of zero entries are the null links. Therefore, we are still able to estimate the missing and likely existing links even the nonzero entries is much less than that being required in Eq. (\ref{eq_lr_inequality}).

When the network is very sparse with high rank or severely corrupted, it is impossible to recover that matrix. However, even in case that the rank of the adjacency matrix is low, the matrix cannot be recovered if the network is extremely sparse \cite{candes2009exact}. This also holds for link prediction, if the network is too sparse which means there are only few connections in a relatively large network, then one cannot correctly predict the unobserved links.

In the matrix completion problem, one supposes that the location of the missing entries are known \cite{candes2009exact}, but usually we have no idea where the locations of the missing entries are. This is also true in the link prediction because only a small part of zero entries of the adjacency matrix are the missing links, and the rest are nonexisting links. Obviously, matrix completion can be treated as a special case of low-rank matrix recovery. Robust PCA is capable of recovering low-rank matrix in the presence of noise as defined in Eq. (\ref{eq_lr}).

\small{
\begin{table*}[ht]
\caption{The prediction precision on the four real weighted networks in which the probe set contains $10\%$ of total connections.}
\centering
\begin{tabular}{*{15}{l}}
\hline
\hline                        
Networks \ & CN \ & AA \  &RA \ & WCN \ & WAA \ & WRA \ & rWCN  \ &  rWAA \ & rWRA \ & LR \ \\ [0.5ex]
\hline
USAir         & 0.349 & 0.369 & \textbf{0.438} & 0.298 & 0.337 & 0.355 & 0.312 & 0.347 & 0.387 & 0.388 \\
C.elegans     & 0.082 & 0.100 & 0.100 & 0.102 & 0.105 & 0.104 & 0.099 & 0.103 & 0.100 & \textbf{0.130} \\
Food web 2   & 0.159 & 0.173 & 0.182 & 0.218 & 0.218 & 0.227 & 0.148 & 0.166 & 0.168 & \textbf{0.345} \\

Football      & 0.133 & 0.133 & 0.150 & 0.067 & 0.133 & 0.133 & 0.083 & 0.117 & 0.117 & \textbf{0.300} \\ [1ex]       

\hline 
\end{tabular}
\label{table:pre_result_weighted}
\end{table*}
}
\begin{figure*}[ht]
\centering
	\includegraphics[width=0.93\textwidth]{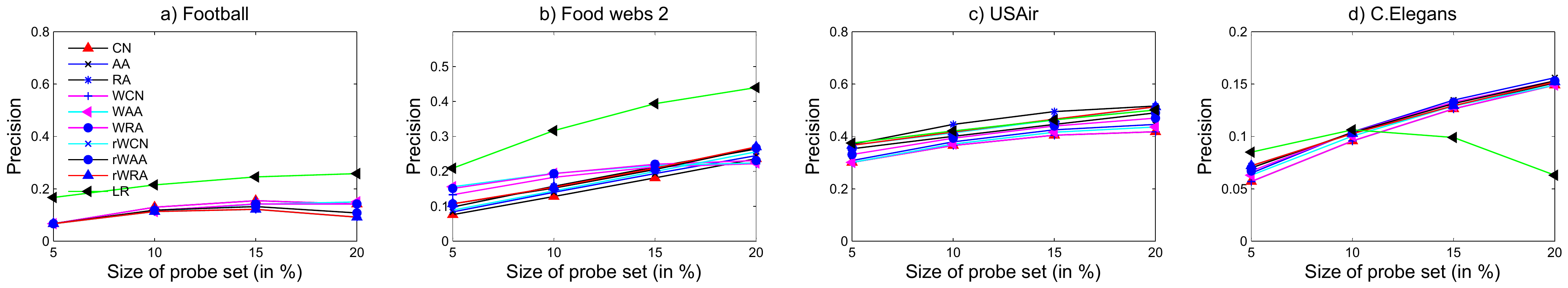}
    \caption{The precision values on the four real weighted networks for different sizes of probe sets.}
\label{fig_probe_w}
\end{figure*}

\section{Simulation}

We implement our matrix completion-based link prediction algorithm and the baseline local similarity-based algorithms in the twelve real networks including eight unweighted and four weighted networks. These networks are
i) Jazz \cite{gleiser2003community} --jazz musician network, link denotes the relationship between two persons if they used to play together in the same band at least once;
ii) Yeast \cite{von2002comparative} --a network of protein-protein interaction;
iii) Political blogs \cite{ackland2005mapping} --a network of hyperlinks between weblogs on US politics;
iv) Hamster \cite{hamster} --a friendship network of users of the web site hamsterer.com;
v) Router  \cite{spring2004measuring}--the router-level topology of the Internet;
vi) Food web 1 \cite{ulanowicz1998FloridaNetwork}--the network of predator-prey interactions in Florida Bay in dry season;
vii) World trade \cite{de2011exploratory}--the network of miscellaneous manufactures of metal among 80 countries in 1994;
viii) Contact \cite{kunegis2013konect} --a contact network between people measured by carried wireless devices, where a vertex represents a person, and a link between two persons shows that there is at least one contact between them;
ix) USAir \cite{Batageli} --the air transportation network of airports;
x) C. elegans \cite{watts1998collective} --the neural network of worm;
xi) Food web 2 \cite{ulanowicz1998network}--the network presenting the predator-prey interactions of Everglades Graminoids in wet season;
and xii) Football \cite{girvan2002community} --the network of American football games consisting of Division IA colleges during the regular season Fall in 2000.
The topology statistics of the twelve networks are shown in Table \ref{table:topology}.

To test the performance of the proposed model, we compare the precision values with six popular unweigted local similarity-based algorithms, e.g., Common Neighbor (CN) \cite{newman2001clustering}, Adamic-Adar (AA) \cite{adamic2003social}, Resource Allocation (RA) \cite{zhou2009predicting}, local community paradigm including CAR, CAA, and CRA \cite{cannistraci2013link}. We call our method as low rank (LR) method, which outperforms the traditional algorithms on those eight unweighted networks. The detailed results are shown in Table \ref{table:pre_result_unweighted}. The precisions on router network computed from all the algorithms are very low as the network is very sparse, i.e., the available information is too limited. The traditional algorithms do not perform well on yeast, hamster and bay dry (food web 2) networks, while LR performs much better. LR performs comparatively better than the others on jazz, public blogs, router and world trade.

Moreover, to show that the proposed method can also deal with weighted network, we compare LR with other six weighted-based algorithms, namely WCN (weighted CN), WAA, WRA \cite{murata2007link}, rWCN (reliable weighted CN), rWAA and rWRA \cite{zhao2015prediction}.
Whenever the link weights become 1, WCN, WAA and WRA are equivalent to CN, AA and RA, respectively. Moreover, WCN, WAA and WRA take the sum of the neighbors' weights into consideration, while rWCN, rWAA and rWRA take the multiplication instead (see details in Refs. \cite{murata2007link,zhao2015prediction,lu2010link}). As shown in Table \ref{table:pre_result_weighted}, the proposed method, in overall, outperforms the others on C.elegans, food web 2 and football network. However, for USAir network, RA performs the best following by LR and rWRA. The predictions on C.elegans fall down when the probe sets are over $15\%$ resulting from sparse and high-rank properties. Excluding rWRA and LR, traditional unweighted-based algorithms outperform weighted-based algorithms on USAir, and it is also reported in \cite{murata2007link}, which may be resulted from the weak ties \cite{lu2010link} effects.

From the empirical simulation, we can see that the local similarity-based algorithms do not perform well on highly dense networks, e.g., food web 1 and 2, hamster, yeast and football, while LR can generate better predictions. To test the sensitivity of the proposed method toward the density of network, we compare the results based on different size of probes set inf Fig. \ref{fig_probe_uw}, Fig. \ref{fig_probe_w}, respectively.

\section{Conclusion and discussion}
In this work, we adopt robust principal component analysis to solve link prediction problem. The adjacency matrix of of the target network is decomposed into low-rank matrix which can be regarded as the backbone of network containing the true links and sparse matrix consisting the corrupted or spurious links in the network. Link prediction, actually, can be regarded as matrix completion problem from corrupted or incomplete adjacency matrix. By solving the optimization problem, we obtain the low-rank matrix which later on plays a role as score matrix illustrating the possible connectivity between each pair of vertices.

Under the assumption that the target network is sufficiently dense and connected, the low-rank recovery technique performs better comparing to other traditional algorithms. When the network gets very dense (see e.g., food webs, Hamster, and Football), the local similarity-based methods are poor while LR performs very well, indicating that the low-rank matrix recovery can well utilize the dense information in adjacency matrix while the local similarity indices cannot. All of the networks we employ in this paper are undirected, however, we strongly believe that the framework can be extended to deal with directed networks.

One can observe that whenever the network is extremely sparse such as router as shown in Table \ref{table:topology}, LR and all the traditional algorithms do not perform well. One might argue that political blogs is not as sparse as router, while LR is less effective (the precision is low). It is because the rank of adjacency matrix is very large. On the other hand, even the ranks of jazz, yeast, food web 1, world trade and food web 2 are also very large, the densities of those networks are also large, then LR still performs well. One can conclude that LR performs well on the network that is sufficiently dense and the rank of the adjacency matrix is not too large, that is to say, LR prefers higher $\mathrm{D}$ and smaller $\tau$.

\begin{acknowledgements}
The authors thank Jian Gao and Qian-Ming Zhang for useful discussions. This work was partially supported by the National Natural Science Foundation of China (NNSFC) under Grant Nos. 61433014.
\end{acknowledgements}

\bibliographystyle{unsrt}
\bibliography{lr_link_pre}

\end{document}